\documentclass[10pt,letterpaper,twocolumn]{article} 

\usepackage{ol2}
\usepackage[draft,implicit=false]{hyperref}
\usepackage[utf8]{inputenc}
\usepackage{amsmath}
\usepackage[square,numbers]{natbib} 
\usepackage{import}

\newcommand{\txtcircled}[1]{\protect\raisebox{.5pt}{\textcircled{\protect\raisebox{-.9pt}{#1}}}}

\begin{document}

\twocolumn[ 

\title{Analysis and filtering of phase noise in an optical frequency comb \newline at the quantum limit to improve timing measurements}

\author{Roman Schmeissner$^1$, Valerian Thiel$^1$, Clément Jacquard$^1$, Claude Fabre$^1$ and Nicolas Treps$^{1,*}$}

\address{
$^1$Laboratoire Kastler Brossel, Université Pierre et Marie Curie, CNRS, ENS, 4 Place Jussieu, 75252 Paris Cedex
$^*$Corresponding author: roman.schmeissner@spectro.jussieu.fr
}

\begin{abstract}It is shown that the sensitivity of a highly sensitive homodyne timing measurement scheme with femtosecond (\textit{fs}) lasers \citep{B.Lamine2008} is limited by carrier-envelope-phase (CEO) noise. We describe the use of a broadband passive cavity to analyze the phase noise of a Ti:Sapph oscillator relative to the standard quantum limit. This cavity also filters lowest levels of classical noise at sidebands above $100$\,kHz detection frequency. Leading to quantum limited CEO-phase noise at $\mu$s-timescales, it can improve the sensitivity of the homodyne pulse timing measurement by 2 orders of magnitude.
\end{abstract}

\ocis{140.0140, 270.0270}

] 

\textbf{Introduction.} Femtosecond optical frequency combs have revolutionized optical metrology \citep{udem2002optical} given their intrinsic frequency comb structure and the ability to measure and lock the phases of these frequencies. Recent experiments have demonstrated that they can be extended to massively parallel optical spectroscopy, using a dual comb configuration and an optical cavity to enhance the recorded signal \citep{Bernhardt2010}. Their dual time and frequency structure also make them ideal candidate for ranging or clock synchronization \citep{Cui2008}. This measurement has been proven to be optimal when used in a homodyne configuration with a pulse shaped reference beam \citep{B.Lamine2008}. The underlying concept of projection on the temporal mode carrying the information can be extended to general parameter estimation  \citep{Jian2012real}.

In this context of very high sensitivity metrology, measurements are limited by the low-level intrinsic noise of the laser, being of classical or quantum nature. Although the noise of Ti:Sapph oscillators has been characterized extensively relative to the carrier \citep{sutyrin2012frequency}, no data are available relative to the quantum limit. We demonstrate here that it is possible to efficiently measure and even filter noise at such low levels. To this aim we combine shot noise resolving intensity noise detection and the filtering properties of a passive cavity.

The paper is organized in the following way: The carrier-envelope-offset (CEO) phase noise, and for completeness the amplitude fluctuations of a commercial Ti:Sapph laser are determined relative to their common quantum limit. A broadband passive optical cavity is then proposed to filter the remaining fluctuations of the CEO phase. Besides filtering, it allows the detection of phase noise down to the quantum limit. The resulting realistic sensitivities for the homodyne timing measurement scheme \citep{B.Lamine2008} are discussed at the end of this paper.

\begin{figure}[t]
\includegraphics[width=8.3cm]{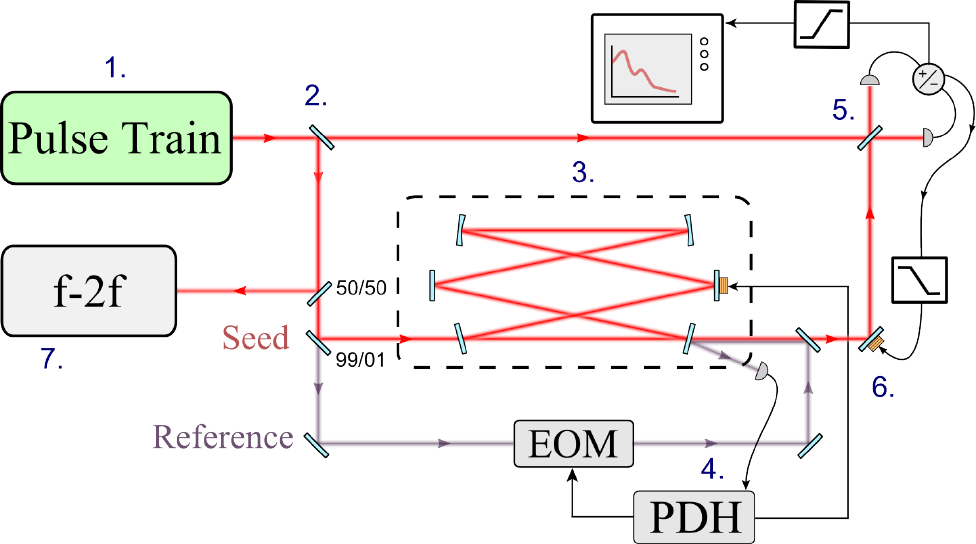}
 \caption{Experimental scheme. 1. Mode-locked Ti-Sapph oscillator, 2. selection of measurement type using a flip 50/50 beamsplitter: amplitude noise measurement if released, interference measurement if set, 3. passive cavity in a mbar vacuum chamber, 4. PDH-locking scheme, 5. balanced (homodyne) detection, 6. lock of the relative phase of the interfering beams, 7. f-2f interferometer} \label{Fig:setup}
\vspace{-0.3cm}
\end{figure}

\textbf{Theoretical concept.} 
We consider a train of $fs$ pulses generated by a commercial mode-locked Ti:Sapph oscillator (Fig.\ref{Fig:setup}). It can be described as a superposition of equally spaced monochromatic modes \citep{weiner2011ultrafast} of frequencies \mbox{$\omega_m = \omega_{\rm CEO}+m\cdot \omega_{\rm rep}$}, where $\omega_{\rm rep}$ is the repetition rate and \mbox{$\omega_{\rm ceo}=2\pi f_{\rm CEO}$} the CEO frequency. But one can also give a time representation of this pulse train. Introducing the light-cone variable \mbox{$u=t-z/c$}, one can write the positive frequency component of the electric field as \mbox{$\mathbf{E}^{(+)}(u)=\mathcal{E}\sqrt{N}\sum_k v(u-kT_{\rm rep})e^{ik\theta_{\rm CEO}}e^{i\omega_0t}$}, where $\omega_0$ is the carrier frequency, $\theta_{\rm CEO}$ the CEO phase, $T_{\rm rep}$ the pulse to pulse time interval, $N$ the photon number in a single pulse, $\mathcal{E}$ a normalization constant and $v(u)$ is a normalized single pulse mode (non zero in interval \mbox{$[0,T_{\rm rep}]$}).

This paper studies properties of frequency comb noise and its filtering in view of possible application to ultra sensitive pulse-timing measurements, such as those introduced in \citep{B.Lamine2008}. These measurements rely on the homodyne detection of a pulse train in mode $v(u)$ with a local oscillator in a superposition of two modes that are proportional to I: $v(u)$ and II: $dv(u)/du$. Using this formalism, the minimum resolvable timing jitter of a fs-pulse train is at 1\,Hz resolution bandwidth and an analysis frequency $f$:\vspace{-0.2cm}
\begin{equation}
\Delta u_{\text{\rm min}}(f) =\frac{1}{2\sqrt{N}}\frac{\left[\omega_{0}^{2} \sigma_{\rm P,\text{I}}^{2}(f)+(\Delta\omega)^2 \sigma_{Q,\text{II}}^2(f)\right]^{1/2}}{\omega_{0}^{2}+(\Delta\omega)^{2}} \label{Eq:1}
\end{equation}
It is a function of the optical bandwidth $\Delta\omega$ of the signal. The normalized variances $\sigma^2(f)$ of the field quadratures P (phase) and Q (amplitude) of the modes I and II are related to: $\sigma^2_{\rm P,\text{I}}$ the CEO-phase of the signal comb, $\sigma^2_{\rm Q,\text{II}}$ the amplitude quadrature of a mode corresponding to a time-of-flight (TOF) measurement.  Both are equal to unity when the light source noise is simply vacuum quantum noise. This level is called the Standard Quantum Limit (SQL). This theoretical result of \citep{B.Lamine2008} can be linked to the experimentally accessible, single sideband (SSB) noise power spectral densities (PSD) $S_{\rm P/Q}$ of the quadratures P and Q of the CEO-phase (I) and the repetition rate (II):
\begin{equation}
\sigma^2_{\rm P_{I}/Q_{II}}(f)=\frac{\sigma^2_{\rm P_{I}/Q_{II}}(f)}{\sigma^2_{\rm SQL}}=\frac{S_{\rm P_{CEO}/Q_{rep}}(f)}{S_{\rm SQL}} \label{Eq:3}
\end{equation}

$S_{\rm SQL}$ is the common quantum limited PSD in unities of dBc/Hz. Given by $S_{\rm SQL}=2\hbar\omega_0/P$, it is a function of the average total comb power $P$. The measurement of both CEO phase noise and TOF jitter is discussed below.

\textbf{Optical setup.} The setup to access and manipulate the noise properties of an optical frequency comb is drafted in Fig.\ref{Fig:setup}. It consists of a commercial Femtolasers\textregistered\ Ti:Sa oscillator emitting $25$-fs pulses at $156$\,MHz repetition rate, centered at $800$\,nm with an average power of 1\,W. A passive, impedance matched cavity, synchronous with the pump laser is placed on one arm of a Mach Zehnder-like configuration, closed by a balanced homodyne detection. A Menlo Systems\textregistered\ f-2f interferometer detects the CEO frequency and its fluctuations after coherent spectral broadening. It allows to lock the Ti:Sapph CEO frequency with a typical bandwidth of less than $20$\,kHz.

\begin{figure}[t]
\centerline{
\includegraphics[width=8.3cm]{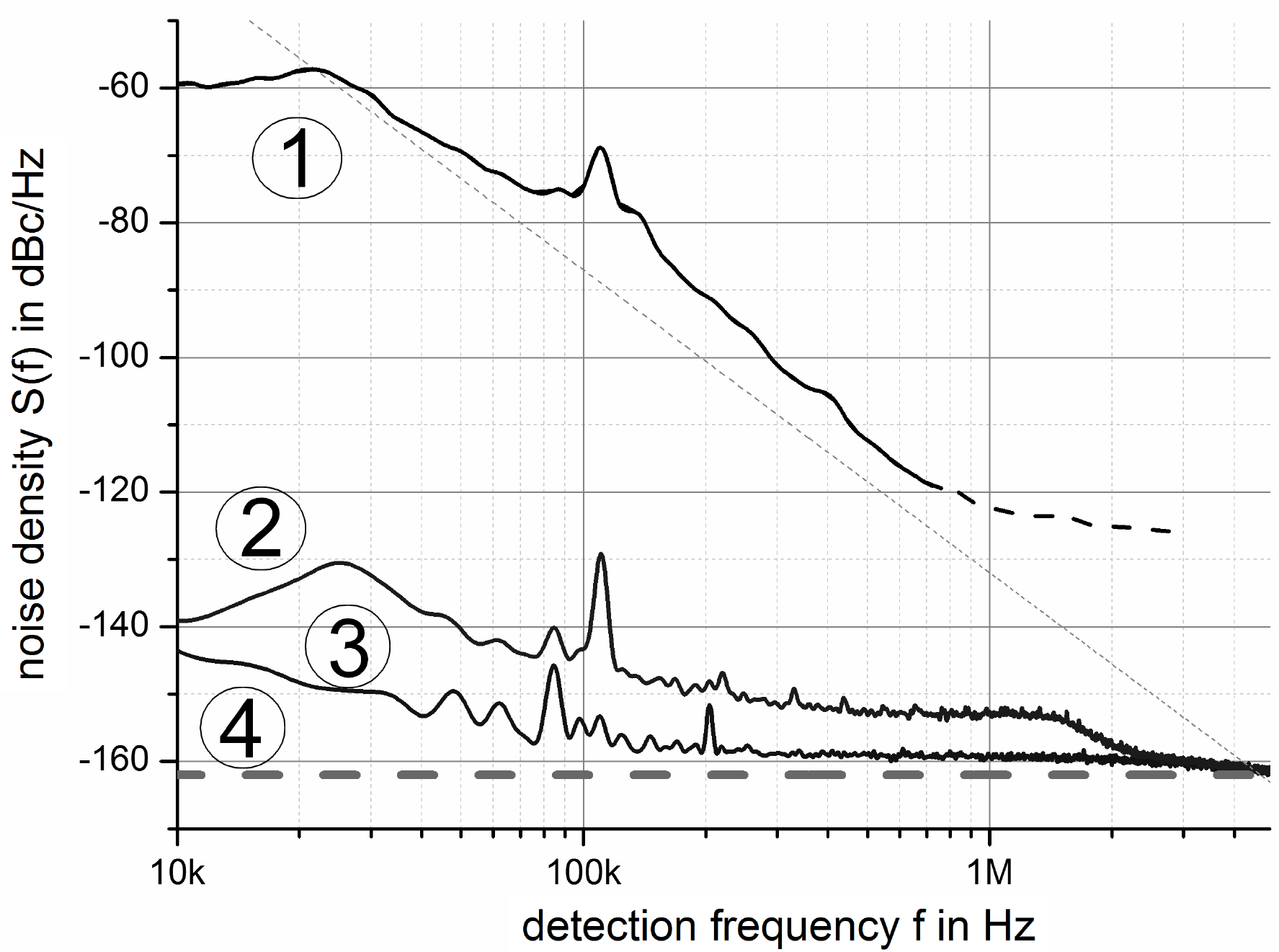}}
 \caption{Noise of a free-running, mode locked Ti:Sa oscillator. \txtcircled{1} Measured CEO phase-noise SSB power spectral density. The noise floor of detection is at \mbox{-125\,dBc}, excess noise from detection is present up to \mbox{-120\,dBc}. \txtcircled{2} Ti:Sa relative intensity noise (RIN) detected with 20-kHz high-pass filters. \txtcircled{3} SQL for both intensity and phase noise at 8\,mW detected signal. \txtcircled{4} calculated SQL} \label{Fig:noise}
\vspace{-0.15cm}
\end{figure}

\textbf{Noise properties of the free-running laser.}
Intensity noise is measured using only the balanced detection. Together with the CEO-phase noise, those data are shown in Fig.\ref{Fig:noise}. When compared to the common SQL for a signal of 8\,mW, the relative intensity noise (RIN) reaches the quantum limit above 3\,MHz. The tiny relaxation oscillation peak at 1.5\,MHz depends on alignment and output power. It has been minimized for this measurement. The locked $f_{\rm CEO}$ can be considered as free running above the lock-resonance at 30\,kHz. The line at 100\,kHz results from the relaxation oscillation of the pumping Verdi\textregistered\ laser, which is also present in the amplitude noise. The CEO-phase noise follows an approximated $f^{-4.5}$ distribution over more than one decade until the noise level of detection is reached at 700\,kHz. It is set by phase-excess noise from the avalanche photodiode detecting the f-2f beating. The $f^{-4.5}$ behavior corresponds to the theoretical prediction of \citep{H.A.Haus1993} to $f^{-4}$ and to an additional term from the coupling of amplitude to phase noise. The levels observed are more than 60\,dB above the repetition rate phase noise. The latter was observed to follow a $f^{-4}$ dependence down to 10\,kHz sideband frequency (data not shown). The noise of the amplitude quadrature of the TOF mode (II) is consequently negligible in Eq.\eqref{Eq:1}. The first term of the equation, corresponding to a phase measurement, is the dominant one. From Eq.\eqref{Eq:1} and Eq.\eqref{Eq:3}, the sensitivity of the homodyne timing measurement \citep{B.Lamine2008} can be approximated to:
\begin{align}
\Delta u_{\text{\rm min}}(f) \approx\frac{1}{2\sqrt{\tau}}\frac{\sqrt{S_{\rm CEO}(f)}}{\omega_0} \label{Eq:2}. 
\end{align}
This expression scales as $1/\sqrt{\tau}$ (with $\tau$ being the measurement time) and with the square root of the spectral density of CEO-phase fluctuations $S_{\rm CEO}$. 

\begin{figure}[t]
\centerline{
\includegraphics[width=7.8cm]{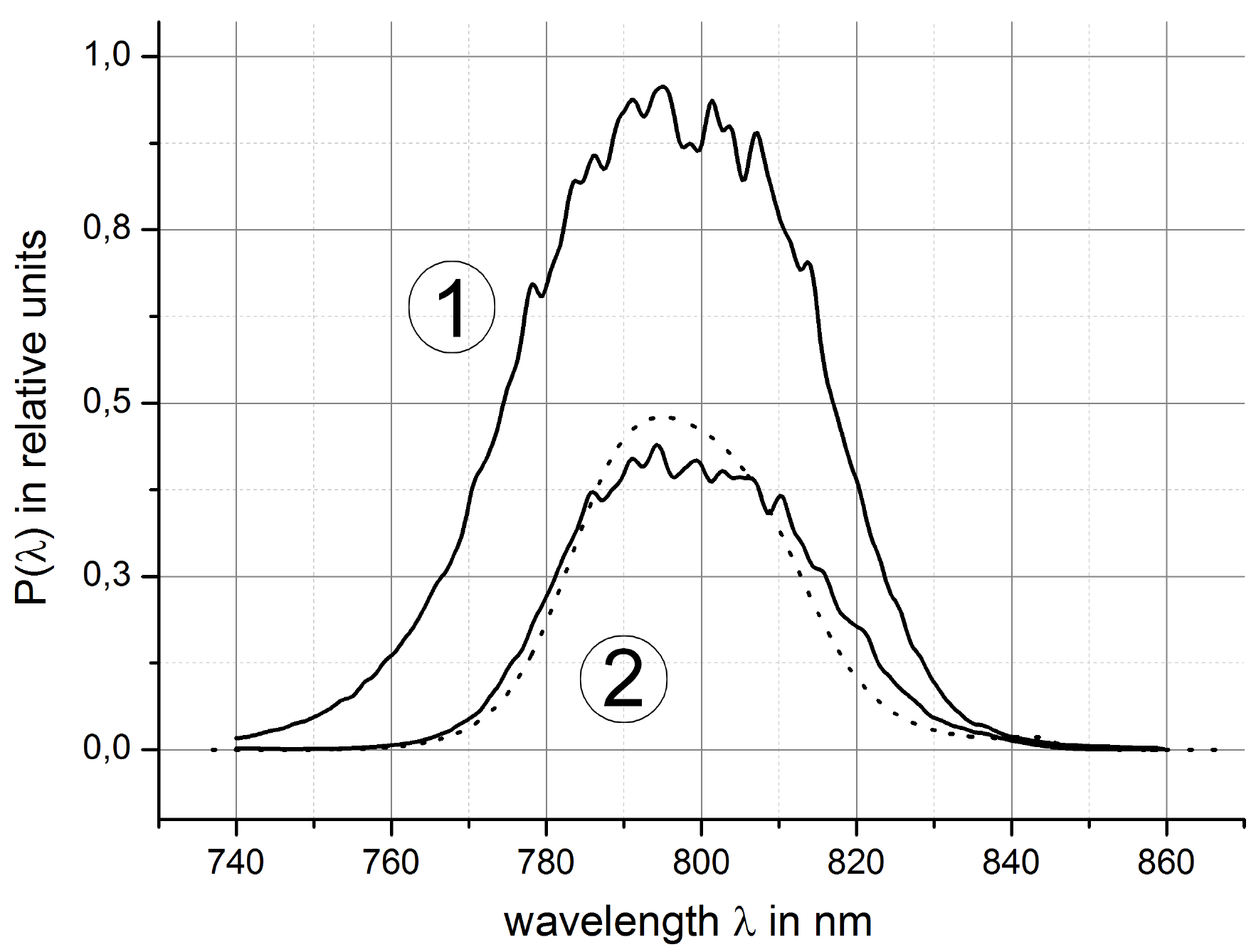}}
 \caption{Cavity transmission spectrum showing broadband simultaneous resonance. \txtcircled{1} A typical seed spectrum of the mode-locked Ti:Sa oscillator. \txtcircled{2} Transmitted spectrum at \mbox{$p$=68mbar} and \mbox{$f_{\rm CEO}=45$\,MHz}, \textit{dotted} simulation at \mbox{$p$=30\,mbar}. At optimized conditions, $~38\%$ of the seed power are transmitted.} \label{Fig:tspectra}
\vspace{-0.2cm} 
\end{figure}

The sensitivity is therefore limited by CEO-phase noise. Active locking schemes typically have a bandwidth of a few kHz. Sidebands above $100$\,kHz and levels close to the SQL are difficult to reach\citep{vernaleken2012carrier}. The next section shows that a passive cavity can filter phase noise above \mbox{$100\,\rm kHz$} and down to the SQL. \\
\textbf{A filtering cavity.} A transmissive optical cavity is a well known 2nd-order low-pass filter acting on both phase and amplitude noise of the input field \citep{hald2005efficient}. The \mbox{3-dB} cutoff frequency \mbox{$f_c=c/(F\cdot L)$} is determined by the speed of light $c$, the cavity finesse $F$, and its \mbox{length $L$}.
We have developed an impedance-matched passive cavity in a bow-tie geometry consisting of $6$ zero-dispersion mirrors and contained in a low-vacuum chamber. Its integration into the experimental setup is depicted in Fig.\ref{Fig:setup}. Residual dispersion is compensated by laboratory air. The constant pressure of $50\pm30$\,mbar is chosen depending on the required spectral shape of transmission. Input and output couplers have equal reflectivities of $99.82\%$ and the 4 other mirrors are highly reflective. The cavity length is set to the same value as the femtosecond oscillator ($1.92$\,m). A Pound-Drever-Hall scheme is used to lock the cavity on resonance with the seed pulse train. To avoid the modulated reference to appear in the detected signal, we use  a counter-propagating reference-beam to generate the error signal. The CEO-phase of the laser is locked to match the resonance frequencies of the filtering cavity. When all locks are running, the 45-nm FWHM spectrum generated by the Ti:Sapph oscillator is entirely (35\,nm FWHM) transmitted through the cavity, see Fig.\ref{Fig:tspectra}. Comparing input and output spectra in Fig.\ref{Fig:tspectra}, we estimate that the residual dispersion of the cavity is less than $2\,\text{fs}^{2}$ over the covered spectrum \citep{Schliesser2006}. The simulations agree  with the observed spectra for a slightly different pressure. The difference may be attributed to uncertainties in the mirror properties. Concerning noise filtering, we measured an effective finesse of \mbox{$F\approx1200$}. The 3-dB cutoff frequency of this cavity was measured at \mbox{$f_c\approx 130$\,kHz}.

\textbf{Measurement of relative phase noise.} To quantify the effect of phase noise filtering by the passive cavity, relative phase noise between a filtered and an unfiltered beam have to be measured down to the SQL. A shot-noise-resolving, balanced homodyne detection \citep{schumaker1984noise} can do so. Its implementation is shown in Fig.\ref{Fig:setup}. It measures phase noise differences before and after the filtering cavity. Two incident fields \mbox{$E_{i}=A_{i}e^{i\phi_{i}}, i=\rm 1,2$} interfere. Locking on the phase quadrature \mbox{$\phi=\phi_1-\phi_2=\pi/2$} leads to the beating signal \mbox{$H=A_{\rm 1}A_{\rm 2}\sin(\delta\phi)$} where $\delta\phi$ are the zero mean fluctuations of the relative phase of the interfering fields. This signal efficiently converts phase- to amplitude fluctuations. If significant amplitude noise is present, one of both beams is strongly attenuated. This turns the scheme into a homodyne detection of relative phase noise. An interference of a bright \textit{local oscillator} (LO) and an attenuated \textit{signal} is less sensitive to amplitude fluctuations. 

Assuming perfect quantum efficiency of the photodetectors, setting the elementary charge to one and neglecting higher order terms, the mean squared homodyne signal reads \citep{haus2000electromagnetic}:
\begin{align}
\mathbf{S}=\frac{\langle\delta H^2\rangle}{(\hbar\omega_0)^2}\cong \frac{1}{(\hbar \omega_0)^2} A_{1}^2A_2^2\langle\delta\phi^2\rangle.
\label{Eq:4}
\end{align}
It detects relative phase fluctuations $\delta \phi$ of the interfering beams. For a given, over-all detected intensity \mbox{$A_0^2=A_{1}^2+A_{2}^2$}, the electronic signal is maximized for equal intensities to \mbox{$\mathbf{S}_{\rm MAX}=A_0^4/(4\hbar \omega_0)^2\cdot\langle\delta\phi^2\rangle$}. This classical signal is always detected relative to the shot noise level \mbox{$\mathbf{N}=A_{0}^2/\hbar\omega_0$}. The detected phase noise can be expressed relative to the carrier in units \mbox{dBc/Hz=(1/2)\,$\cdot$\,rad$^2$/Hz} \citep{R.P.Scott2001}. For a signal to noise ratio \mbox{SNR=$\mathbf{S/N}=1$} and a detected power $P$, the minimal resolvable relative phase noise is
\begin{align}
\langle\delta\phi^2\rangle_{\rm min}=\frac{8\hbar\omega_0}{P}=4S_{\rm SQL}
\end{align}

\begin{figure}[t]
\centerline{
\includegraphics[width=8.3cm]{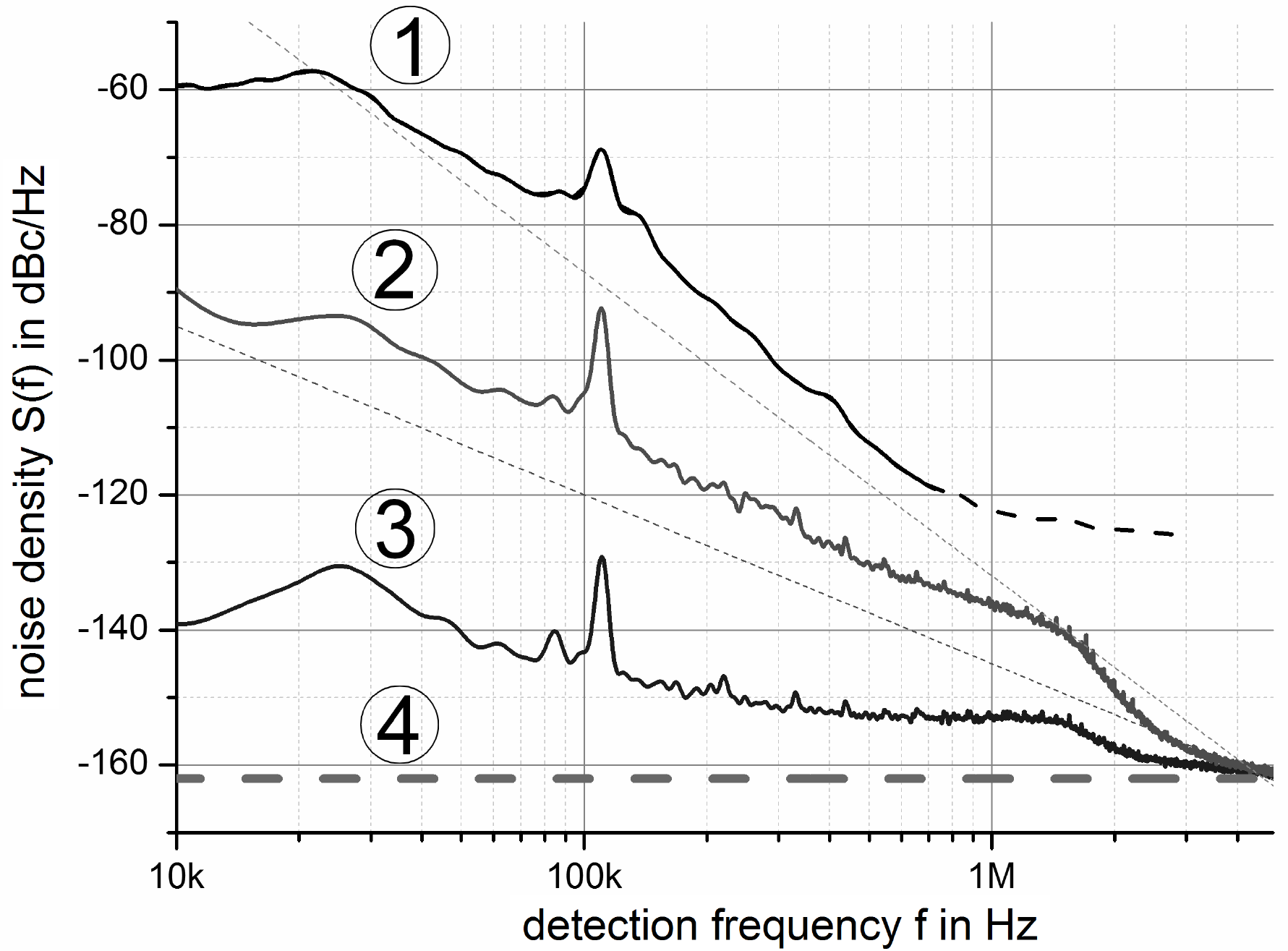}}
\caption{Relative phase noise after filtering with the cavity. \txtcircled{1} SSB power spectral density of CEO phase noise from f-2f measurment, \textit{dotted} $f^{-4.5}$ power law, \txtcircled{2} Phase quadrature of the homodyne beating signal between the Laser (Local Oscillator) and the cavity-output (Signal), \textit{dotted} $f^{-2.5}$ power law, \txtcircled{3} Ti:Sa RIN, \txtcircled{4}\  SQL for 8mW detected signal} \label{Fig:cavlas}
\vspace{-0.0cm}
\end{figure}

Consequently, given sufficiently low amplitude noise, the measurement scheme Fig.\ref{Fig:setup} of relative phase noise has a quantum limited sensitivity. For sufficiently high detection frequencies, the phase noise in the filtered arm becomes negligible. The detected signal is than proportional to the CEO-phase noise $\delta \phi\approx \delta \theta_{CEO}$. With \mbox{$f_c\approx 130$kHz} used here, this relation holds at MHz detection frequencies. In conclusion, absolute levels of CEO-phase noise also become measurable down to the SQL. 

 \textbf{Measurement Data.} The measured homodyning signal is shown in Fig.\ref{Fig:cavlas}, trace \txtcircled{2}. It arises from the interference of the signal from the Ti:Sapph oscillator (LO) with the 10\,dB less intense beam filtered from the cavity (Signal). The signal from the cavity exhibits intensity excess noise arising from noise-quadrature interconversion by the cavity (not shown). Nevertheless, its presence does not change the RF-spectral distribution of the homodyning signal. Classical intensity noise of both the signal and the local oscillator cancel in the balanced measurement configuration and only contribute to higher order terms of the signal $\mathbf{S}$ described in Eq.\eqref{Eq:4}. 

\begin{figure}[t]
\centerline{
\includegraphics[width=8.3cm]{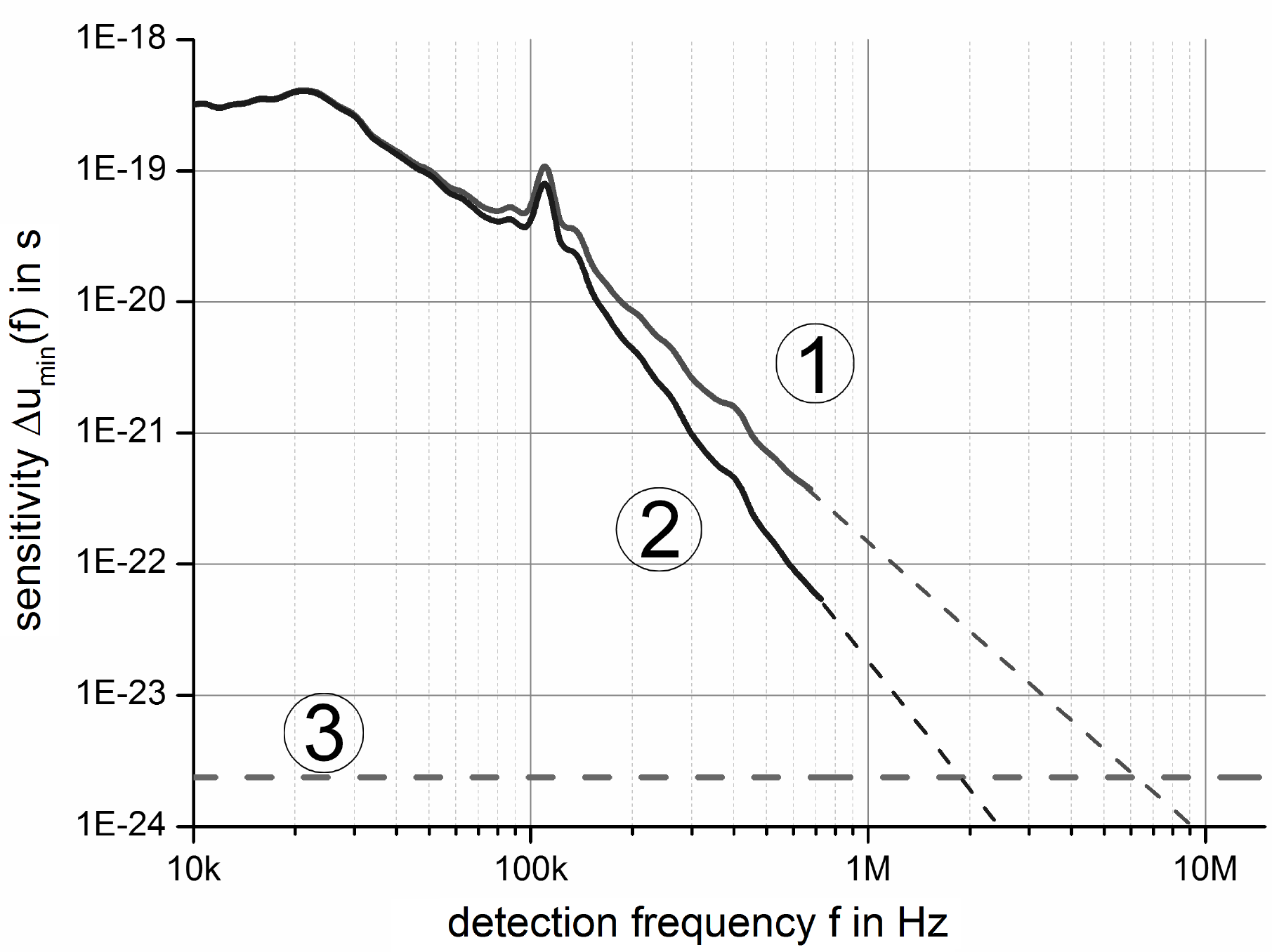}}
 \caption{Predicted realistic sensitivity of a projective timing measurement using mode-locked Ti:Sa lasers \citep{B.Lamine2008}, using an unfiltered \txtcircled{1} and a cavity-filtered \txtcircled{2} oscillator. \textit{dashed lines} extrapolation of available data \txtcircled{3} Quantum limit for phase and amplitude noise at the SQL for 8\,mW detected signal at 1\,s integration time.} \label{Fig:msprecis}
\vspace{-0.2cm} 
\end{figure}

\textbf{Phase noise filtering and detection.} The experimental results are plotted in Fig.\ref{Fig:cavlas}. The interfering beams have a spectral overlap of 94\%. The homodyne signal of the phase quadrature of mode I follows an approximate $f^{-2.5}$ power law over nearly two decades of detection frequencies. Indeed, the CEO-phase noise of the Ti:Sapph oscillator was observed to follow an $f^{-4.5}$ dependence. In addition, the filter-efficiency of the cavity is then given by its transfer function which follows a $f^{2}$ distribution. Consequently, the expected power law for the relative phase noise is the product of both, $f^{-2.5}$. This is equivalent to the measured distribution. It confirms, together with the transmitted optical bandwidth, that cavities can not only filter noise of single optical frequencies \citep{hald2005efficient} but also of entire coherent frequency combs. The measured amplitude noise of the oscillator is neglectable for the measurement above (see Fig.\ref{Fig:cavlas}, trace \txtcircled{3}). Similar to the amplitude noise, the homodyne signal and thus the CEO-phase noise of the oscillator vanishes in shot noise at approximately \mbox{$5$\,MHz} detection frequency (see Fig.\ref{Fig:cavlas}, trace \txtcircled{2}). 

The consequences of the observed phase noise filtering for the timing measurement discussed with Eq.\eqref{Eq:2} are shown in Fig.\ref{Fig:msprecis}. Using the f-2f CEO-phase noise data, the possible precision of the homodyne timing measurement \citep{B.Lamine2008} can be calculated for a filtered or an unfiltered beam. From Fig.\ref{Fig:cavlas} it follows from the interference data that the slope of the CEO-phase noise does not significantly change at microsecond timescales. The achievable measurement precision shown in Fig.\ref{Fig:msprecis} can thus be extrapolated down to the SQL. Using a passive cavity to filter phase noise, the expected sensitivity of the timing measurement could be improved by up to 2 orders of magnitude .

\textbf{Conclusions.} A broadband resonant, passive cavity has been shown to be a tool for filtering and the detection of CEO-phase noise of a \mbox{$45$-nm} FWHM frequency comb. Together with shot noise resolving balanced detection, it is shown that a commercial Ti:Sapph oscillator is quantum limited in amplitude and phase below 5\,MHz detection frequency. Passive filtering of phase noise decreases this frequency. It potentially improves the sensitivity of a pulse-timing measurement scheme by up to two orders of magnitude. 

\textbf{Acknowledgments.} The authors thank Yann Le Coq for helpful discussion. 
The research is supported by the ERC starting grant Frecquam. C.Fabre is
a member of the Institut Universitaire de France.

\bibliographystyle{ol} 
\end{document}